\documentstyle[aps,prl,epsf]{revtex}
\begin{document}
\draft

\title{ Nonequilibrium Magnetization Dynamics of Nickel}

\author{J.Hohlfeld, E. Matthias\\
Institut f\"ur Experimentalphysik, Freie Universit\"at Berlin, 14195 Berlin, Germany}

\author{R. Knorren, K.H. Bennemannn\\
Institut f\"ur Theoretische Physik, Freie Universit\"at Berlin, 14195 Berlin, Germany}

\date{}

\maketitle
\begin{abstract}
  Ultrafast magnetization dynamics of nickel has been studied for
  different degrees of electronic excitation, using pump-probe
  second-harmonic generation with $150\,$fs/$800\,$nm laser pulses of
  various fluences. Information about the electronic and magnetic
  response to laser irradiation is obtained from sums and differences
  of the SHG intensity for opposite magnetization directions. The
  classical M(T)-curve can be reproduced for delay times larger than
  the electron thermalization time of about $280\,$fs, even when
  electrons and lattice have not reached thermal equilibrium. Further
  we show that the transient magnetization reaches its minimum
  $\approx 50\,$fs before electron thermalization is completed.
\end{abstract}


\narrowtext

Ultrafast spin dynamics in ferromagnets is of great interest from both
theoretical and experimental points of view. In particular, the
short-time dynamics of magnetism in transition metals, with many
excited electrons not at equilibrium with the lattice, is a new area
of physics. Such studies are important for developing a theory of
transient magnetization behavior in the subpicosecond range. It seems
that the only experimental data which can guide theoretical analysis
are the ones reported by Beaurepaire et al.  \cite{BMD_96} on
time-resolved demagnetization of Ni induced by femtosecond laser
pulses of $620\,$nm at one specific fluence. The authors utilized the
magneto-optical Kerr effect to detect hysteresis loops for different
time delays between pump and probe pulses. By comparing the
time-dependent remanence with the equilibrium temperature dependence
of magnetization, $M(T)$, they derived the time evolution of the spin
temperature within the framework of the phenomenological
three-temperature model \cite{VBM_91}. Clearly, it is of great
importance to confirm whether or not $M(T)$ can be used to describe
the transient magnetic response to electron excitations in itinerant
ferromagnets and whether there is a time delay between electron
thermalization and magnetization changes.

In this Letter we present time-resolved data on the transient
magnetization measured by pump-probe second-harmonic generation (SHG).
The great advantage of this technique is that it allows to
simultaneously follow electron-temperature relaxation and transient
magnetization, without further need for additional calibration
measurements. This is a consequence of the even and odd contributions
to the nonlinear susceptibility \cite{PWS_89}. The measurements were
carried out for a large variety of pump fluences leading to different
initial electron temperatures. After equilibration of the electron
bath, we find the transient magnetization to be governed by the
electron temperature $T_{e}$ via the classical $M(T)$-curve
\cite{WF_26}.  However, we observe a strong deviation of the data from
the magnetization curve in the short-time range $t < 0.3\,$ps, where
the electron system is in a nonequilibrium state.

When discussing nonequilibrium spin dynamics in ferromagnets, it is
worth noting that relaxation times presented here, where
nonequilibrium electronic states are excited, are not compatible
with the spin-lattice relaxation times, $\tau_{sl}$. The latter
denotes the time it takes for the spin system to adjust to a new
temperature while equilibrium between electron and lattice temperature
is maintained. Hence, the values of $\tau_{sl}=100 \pm 80\,$ps
measured by Vaterlaus {\it et al.} for ferromagnetic Gd
\cite{VBM_91,V_92} and the limits given for Fe \cite{V_92,V_90} and
Ni \cite{A_84} must be considered in a different context.

Experiments were performed at room temperature on polished
polycrystalline nickel (purity 99.90 \%) in air. The samples were
magnetized to saturation with the magnetization direction oriented
perpendicular to the plane of incidence. P-polarized
$150\,$fs/$800\,$nm pump and probe pulses with an intensity ratio of
3:1 for the highest pump fluence were used to excite the electrons and
probe electron temperature and magnetization by detecting the
reflected unpolarized SHG.  Angles of incidence were $20^{\circ}$ for
the pump and $45^{\circ}$ for the probe beams. We do not expect any
interfering effects from the nickel oxide layer --- which is
antiferromagnetic --- since we only measured relative changes in SHG
yield for opposite magnetization directions \cite{BHM_95} as a
function of delay time between pump and probe pulses.

\begin{figure}[htbp]
\epsfxsize=8.cm
\centerline{\epsffile{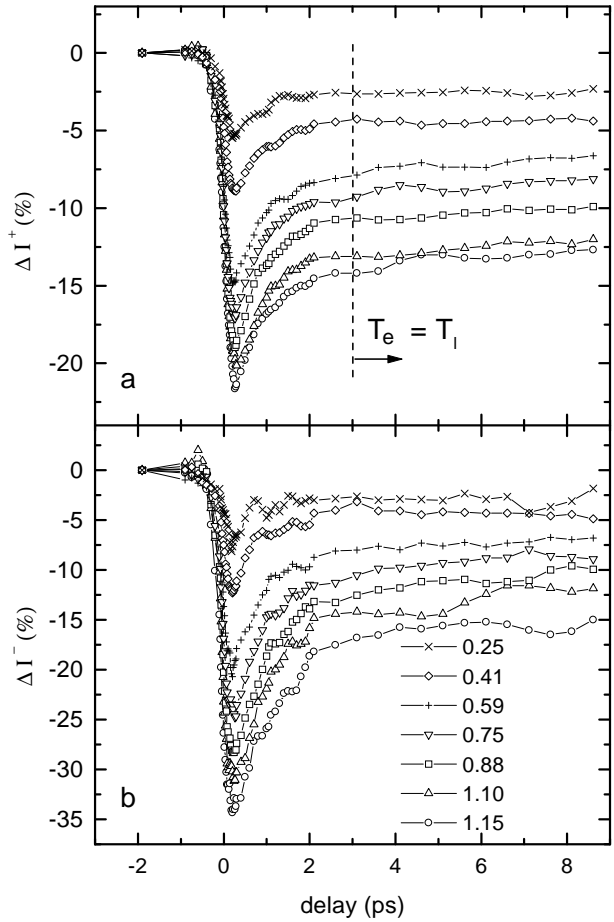}}
\vspace{2ex}
\caption{\label{plusminus} Time dependence of normalized sums
  (a) and differences (b) of SHG yields for opposite magnetization
  directions, as defined in Eq.\,(1). The curves were recorded with
  different relative fluences, calibrated by $1.00 \approx 6$
  mJ/cm$^{2}$.  Constant levels for $t>3\,$ps reflect equilibrium
  between electron and lattice temperatures, $T_{e}$ and $T_{l}$.}
\end{figure}

Typical relaxation curves are shown in Fig.\,\ref{plusminus} for seven
selected pump intensities, although more were measured and included in
Fig.\,3.  Displayed are sums and differences of SHG signals for
opposite magnetization directions $I^{\pm}(t)$, normalized to the SHG
yield $I^{\pm}_{0}$ for the probe pulse alone:
\begin{equation}
\label{ipm}
\Delta I ^{\pm}= \left(I^{\pm}(t)-I^{\pm}_{0}\right)/I^{\pm}_{0} \,\,\, ,
\end{equation}
where $I^{\pm}=I(2\omega,+\vec{M})\pm I(2\omega,-\vec{M})$. The
second-order polarization in the presence of magnetization can be
written in the form \cite{PWS_89}
\begin{equation}
\label{pol}
P_{i}(2\omega)=\sum_{j,k}\left[\chi^{even}_{ijk}(\pm \vec{M})\pm \chi^{odd}_{ijk}(\pm \vec{M})\right] E_{j} E_{k}\,\,\,\, .
\end{equation}
The even tensor elements are almost unaffected by the magnetization, whereas
the odd ones scale linearly with $M$ \cite{PHB_94}. Hence, we can substitute:
\begin{equation}
\label{ceco}
\chi^{even}_{ijk}(M)=\chi^{even}_{0,ijk} \,\,\, , \,\,\, \chi^{odd}_{ijk}(M)=\chi^{odd}_{0,ijk}\cdot |M| \,\,\, .
\end{equation}
From Eq.\,\ref{pol} we obtain the expressions
\begin{eqnarray}
\label{ip}
I^{+}&=&2 \cdot I^{2}_{0}(\omega)\cdot \left[|A \chi^{even}_{0}|^{2}+|B \chi^{odd}_{0} M|^{2}\right] \,\,\, ,\\
\label{im}
I^{-}&=&4 \cdot I^{2}_{0}(\omega)\cdot |A \chi^{even}_{0} B \chi^{odd}_{0} M| \cdot \cos\phi \,\,\, ,
\end{eqnarray} 
where the individual tensor elements are combined to an effective even
and odd contribution denoted by $\chi^{even}_{0}$ and
$\chi^{odd}_{0}$. $A$ and $B$ represent the corresponding effective
Fresnel coefficients for the fundamental and frequency-doubled light
\cite{SMD_87}. The resulting phase of $A$, $B$, $\chi^{even}_{0}$, and
$\chi^{odd}_{0}$ depends on $\vec{M}$ and is given by $\phi$.
Measurements of transient linear reflectivities (not shown here)
verify that the magnitude of the Fresnel factors $A$ and $B$ is nearly
unaffected by $T_{e}$. Also, calculations reveal that the
magnetization-independent factors $\chi^{even}_{0}$ and
$\chi^{odd}_{0}$ vary only a few percent with $T_{e}$ for $T_{e}\le
T_{C}$ ($T_{C}=$Curie temperature). This can be understood by the
large width ($\approx 0.3\,$eV) of vacant d-states in the minority
band \cite{CW_73} in comparison with the maximum electron temperature
of $550\,$K reached with the highest pump fluence.  Thus, we conclude
that the time dependence of $I^{\pm}$ is contained in $M(t)$. Assuming
$M(t)=M(T_{e}(t))$, the classical $M(T)$-curve \cite{WF_26} can be
approximated by $M(T_{e})=M(T_{0})[1- const. \,\,
(T_{e}-T_{0})]^{1/2}$ for the electron-temperature range covered in
our experiment. Here, $T_{0}$ denotes the mean electron temperature
at negative delay. Inserting this into
Eqs.\,(\ref{ipm}), (\ref{ip}), and (\ref{im}) we obtain
\begin{eqnarray}
\label{dip}
\Delta I^{+}(t)&=& const. \cdot [T_{0}-T_{e}(t)] \,\,\, ,\\
\label{dim}
\Delta I^{-}(t)&=& \frac{M\left(T_{e}(t)\right) \cdot \cos\phi}{M(T_{0})}-1 \,\,\, .
\end{eqnarray}
Prerequisite for the validity of these equations is a thermalized
electron distribution. Thus for $t>0.3\,$ps the curves in Fig.\,1a
represent the time evolution of electron temperature, while those in
Fig.\,1b describe the transient magnetization, provided $\phi$ is
constant.  That $\Delta I^{+}$ is indeed proportional to temperature
can be verified by plotting the averaged values for delay times
$>3\,$ps in Fig.\,1a against fluence.  The result is shown in Fig.\,2.
In thermal equilibrium with the lattice, $T_{e}$ scales linearly with
fluence. The slope $m=1$ in Fig.\,2 proves this proportionality in
Eq.\,\ref{dip} and hence the validity of our approximation for
$M(T_{e})$ for longer delay times.

Eqs.\,\ref{dip} and \ref{dim} suggest a graph of $(\Delta I^{-}+1)$
versus $-\Delta I^{+}$ which can be compared with the classical
$M(T)$-curve for nickel \cite{WF_26}. Such plot is shown in Fig.\,3.
In the upper frame [Fig.\,3a], only data for delay times $> 0.3\,$ps
have been analyzed.  Obviously, this plot reproduces within $\pm 5\%$
the equilibrium magnetization curve, indicated by the solid line.  The
good agreement justifies the above made assumptions and demonstrates
that $\phi$ does not depend significantly on $T_{e}(t)$. Furthermore,
it provides an intrinsic electron-temperature calibration for
excitation with various pump intensities. It must be emphasized that
this agreement includes the range between $0.3\,$ps and $3\,$ps where
$T_{e}$ is {\it not} in equilibrium with the lattice temperature [cf.
Fig.\,1]. We also conclude from Fig.\,3a that the electron temperature
governs the magnetization. From the delay time where the points fall
on the magnetization curve we can unambiguously conclude that the
electrons are fully thermalized after $\approx 280\,$fs, when the
minima in Fig.\,1a are reached.

The nonequilibrium situation is demonstrated by incorporating the
data for $t < 0.3\,$ps into the same plot, as illustrated in Fig.\,3b
for three fluences. The strong deviation of the data from the
magnetization curve can be attributed to nonequilibrium electrons,
and, in view of Eq.\,(\ref{im}), to a rapid phase change.  It vanishes
when the thermalization of the electron gas is completed, and then the
data again follow the magnetization curve.

The different behavior of nonequilibrium electrons becomes more
evident in the real-time plot of $\Delta I^{+}$ and $\Delta I^{-}$ in
Fig.\,4, which is a magnified version of the short-time range in
Fig.\,1.  For clarity, only data for three fluences are displayed, but
the same delay between the minima of $\Delta I^{+}$ and $\Delta I^{-}$
is observed for all fluences. The minima of $\Delta I^{+}$ at $(280\pm
30)\,$fs (dotted line in Fig.\,4a) mark the time at which the
thermalization of the electrons is completed and agrees with the value
reported by Beaurepaire et al. \cite{BMD_96}.  The breakdown of
$\Delta I^{-}$ in Fig.\,4b follows the convolution of pump and probe
pulses. For all fluences the minimum is reached about $50\,$fs {\it
  before} the electron temperature is established.  To understand the
faster response of $\Delta I^{-}\propto M$ compared to $\Delta I^{+}$
we notice that for $t<0.3\,$ps Eq.\,(\ref{dip}) is no longer valid and
$\Delta I^{+}\propto |M|^{2}$. This means that $\Delta I^{+}$ reflects
the momentary population difference between minority and majority
spins and, therefore, monitors the development of electron temperature
out of a highly nonequilibrium state \cite{fann}. $\Delta I^{-}$ is
in addition sensitive to the direction of $M$ and samples
preferentially the minority spins of the vacant d-states. Note, that
the spin-orbit coupling is restored faster for minority than for
majority spins \cite{Aesch}, since the width of the vacant d-states
($\approx 0.3\,$eV) is large compared to the Fermi distribution for
$T_{e}\le T_{C}$. The shift in Fig.\,4 and the fact the magnetization
curve is reproduced after thermalization of the electrons prove that
for $t \ge 280\,$fs the electron temperature governs the
magnetization. This is in striking contrast to the results reported by
Beaurepaire et al.  \cite{BMD_96} who observed a $2\,$ps delayed
magnetic response with respect to $T_{e}$. Further investigations are
required to elucidate the cause of these deviating observations, since
too many parameters differ in the experiments reported here and in
\cite{BMD_96}. Note, however, that the response time of itinerant
magnetism is expected to be of the order of $T_{C}^{-1}$,
corresponding to about $80\,$fs for Ni.

Let us return to Fig.\,1 and attempt to understand the large decrease
of $\Delta I^{\pm}$ in terms of the band structure. The fundamental
wavelength of $800\,$nm ($1.55\,$eV) matches well the difference
between the two mains peaks of the Ni minority band \cite{CW_73}, as
sketched in Fig.\,5. Due to the exchange splitting, the upper peak is
unoccupied for minority spins and can resonantly enhance the SHG in
the intermediate state. For majority spins there is no such
enhancement. The filling of the vacant states in the minority d-band
causing the breakdown of $M$ reduces this resonance enhancement which
leads to the fluence-dependent drop of signals in Fig.\,1. The effect
is further amplified by the large matrix elements for d-d transitions
\cite{Hue}. The size of reduction of $\Delta I^{+}$ is determined by
the ratio of odd to even parts of the nonlinear susceptibility and the
corresponding Fresnel factors.  Using the temperature calibration via
Figs.\,2 and 3, one can derive from Eqs.(3) and (4) the ratio $|B
\chi_{0}^{odd}|/|A \chi_{0}^{even}|=0.7\pm 0.1$.

Summarizing, we have shown that pump-probe SHG is a powerful method
for investigating ultrafast spin dynamics in ferromagnets, since it
contains information about both, electron temperature and
magnetization. We applied different pump fluences to study the
electron and spin dynamics for various degrees of excitation.
Independently of pump fluence, the electron thermalization is completed
after $(280\pm30)\,$fs and the magnetization curve \cite{WF_26} is
reproduced from this time on, even when the electrons are {\it not} in
equilibrium with the lattice. For the first time we have observed that
the magnetic response is faster than the electron thermalization,
which may be attributed to different lifetimes of excited majority and
minority electrons. Further investigations of this topic, in
particular frequency-dependent measurements which may uncover
band-structure effects, are desirable.

This work was supported by the Deutsche Forschungsgemeinschaft, Sfb 290.

\newpage

\begin{figure}[htbp]
\epsfxsize=8.cm
\centerline{\epsffile{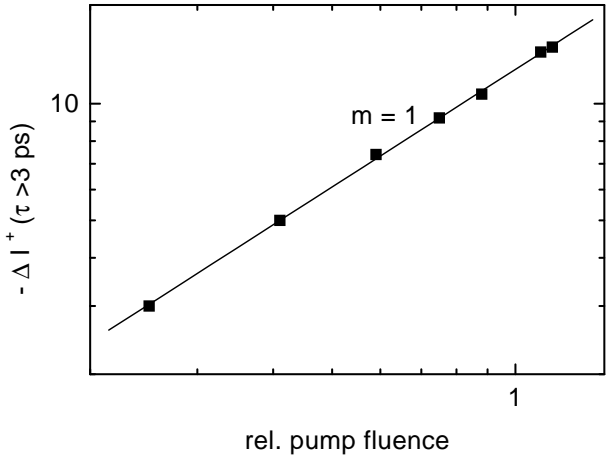}}
\vspace{2ex}
\caption{\label{linear} Normalized sums of SHG yields for opposite 
  magnetization directions, averaged over all values for $t>3\,$ps, as
  a function of relative fluence. The linear slope indicates that
  $\Delta I^{+}$ is proportional to electron temperature [see
  Eq.\,(6)].}
\end{figure}

\begin{figure}[htbp]
\epsfxsize=8.cm
\centerline{\epsffile{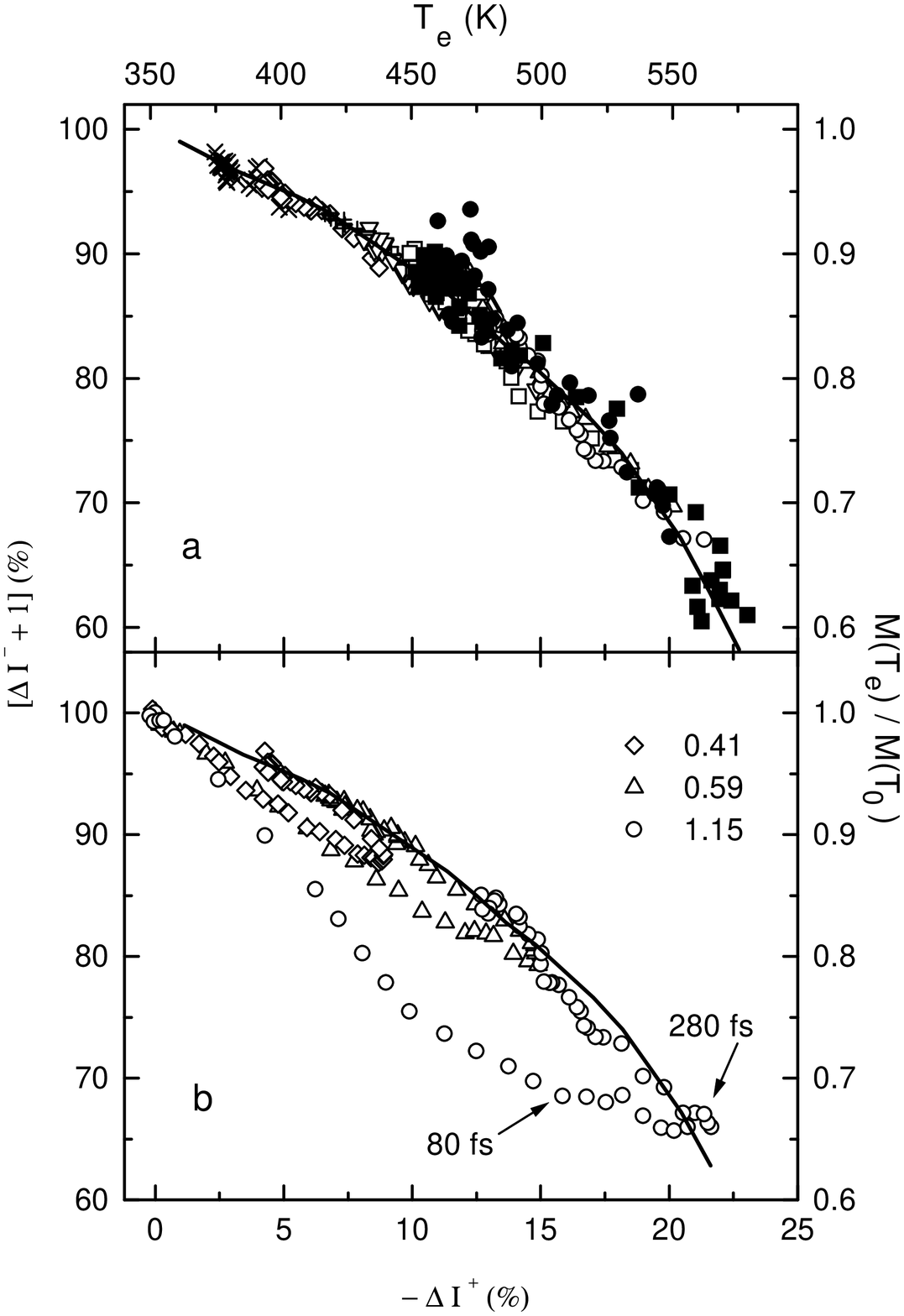}}
\vspace{2ex}
\caption{\label{julius} (a) Comparison of all measured data for 
  $t>0.3\,$ps with the equilibrium magnetization curve of Ref.\,[4],
  represented by the solid line.  (b) Data for three selected fluences
  covering the total time range.  Deviations from the magnetization
  curve point to a nonequilibrium state of the electron and spin
  systems for $t<0.3\,$ps.}
\end{figure}

\begin{figure}[htbp]
\epsfxsize=8.cm
\centerline{\epsffile{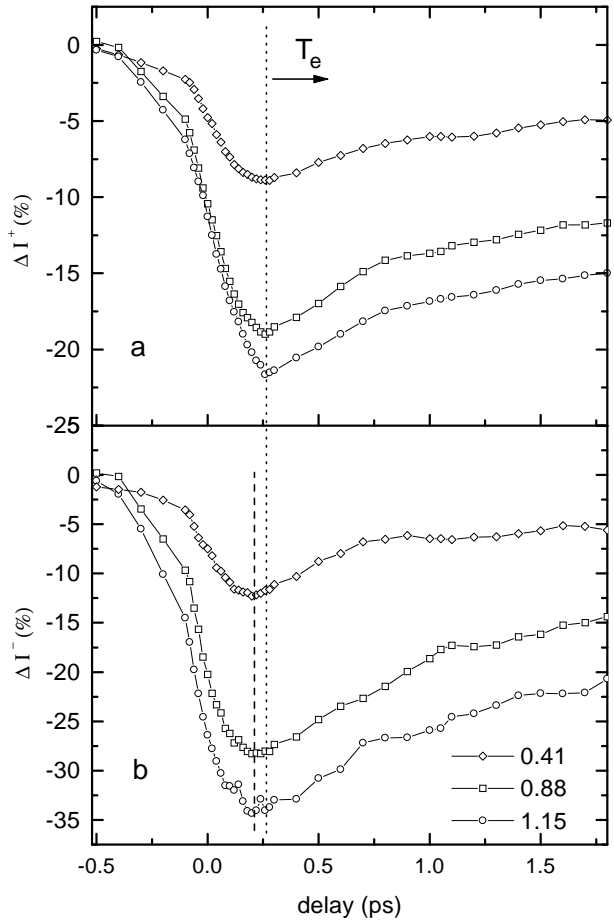}}
\vspace{2ex}
\caption{\label{plusminusmag}  Magnified version of Fig.\,1 for 
  three selected fluences. The minima at $(280\pm30)\,$fs in (a),
  marked by the dotted line, signal the onset of a well defined
  electron temperature. The minima in (b) occur $50\,$fs faster
  (dashed line).}
\end{figure}

\begin{figure}[htbp]
\epsfxsize=8.cm
\centerline{\epsffile{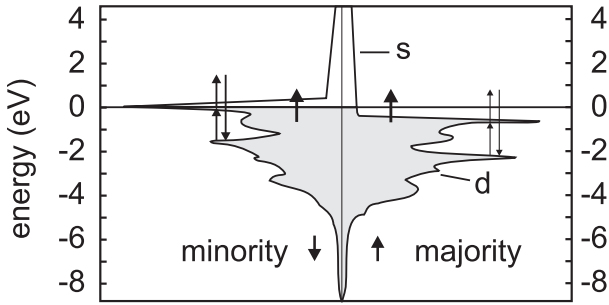}}
\vspace{2ex}
\caption{\label{dos} Density of states for minority and majority spins 
  according to Callaway and Wang [11].  While single-photon excitation
  proceeds with comparable rates from both parts, SHG is resonantly
  enhanced by the unoccupied peak in the minority band.}
\end{figure}

\end{document}